\documentclass[conference]{IEEEtran}
\IEEEoverridecommandlockouts

\usepackage{graphicx}
\usepackage{booktabs}
\usepackage{tabularx}
\usepackage{array}
\usepackage{multirow}
\usepackage{placeins}
\usepackage{makecell}
\newcolumntype{P}[1]{>{\raggedright\arraybackslash}p{#1}}
\begin{document}

\title{MOSS: A Production Framework for Monitoring and Diagnosing Semantic Failures in Search and Recommendation Systems}

\author{
\IEEEauthorblockN{
\begin{tabular}{cc}
Alina Nesen\textsuperscript{1} &
Prathyusha Kanmanth Reddy\textsuperscript{1} \\
David Relyea\textsuperscript{1} &
Rohit Avinash Mujumdar\textsuperscript{2} \\
Samaneh Karimi\textsuperscript{1} &
Bo Shen\textsuperscript{2} \\
\multicolumn{2}{c}{Jordan Hashemi\textsuperscript{1}}
\end{tabular}
}
\IEEEauthorblockA{\textit{Data Sciences, Target Corporation}\\
\textsuperscript{1}Minneapolis, MN, USA \qquad
\textsuperscript{2}Sunnyvale, CA, USA\\
\{alina.nesen, prathyusha.kanmanthreddy, david.relyea,\\
rohitavinash.mujumdar, samaneh.karimi, bo.shen,\\
jordan.hashemi\}@target.com}
\thanks{A prior version of this work was presented at the non-archival
Unified Search and Recommendation Workshop (USRW), co-located with ACM
RecSys 2026. This work has been submitted to the IEEE for possible
publication. Copyright may be transferred without notice, after which
this version may no longer be accessible.}
}

\maketitle

\begin{abstract}
Detecting semantic failures in production recommendation and search systems is hard: behavioral metrics lag by days, manual review does not scale, and most LLM evaluation frameworks are designed for offline, fixed datasets rather than continuous operational monitoring. We present Model Sanity Service (MOSS), a production framework that addresses this operational gap by embedding LLM-based relevance scoring into a recurring monitoring loop with targeted sampling, strategy-aware prompting, structured alerting, and diagnostic follow-up prompting that categorizes failures into actionable root causes.

MOSS has been running in production at a large U.S. retailer for over six months across recommendation, search, and offer surfaces, and has surfaced two confirmed serving defects. For one of these, a backfill blending defect, MOSS quantified the impact of the proposed fix offline as a 76.4\% relative reduction in alert rate. On a 498-pair human-annotated recommendation benchmark, frontier LLM judges achieve higher F1 and AUC point estimates than non-LLM baselines (internal production embeddings and SBERT), with the largest gains on strategy-dependent tasks requiring intent-level relational reasoning. The same judge, score scale, and threshold transfer to a 466-pair search benchmark spanning 21 queries with only a surface-specific prompt adaptation, achieving strong F1 and AUC without retuning. These results show that a single LLM monitoring framework can provide useful, actionable quality signals across multiple product discovery surfaces in a large-scale e-commerce setting.
\end{abstract}


\section{Introduction}
Modern search and recommendation systems are typically evaluated using lagging behavioral metrics such as click-through rate and conversion, which delay detection of semantic failures and provide limited insight into their root causes \cite{zangerle2022survey}. In practice, issues such as category mismatches, intent violations, or degraded relevance may persist until they impact users, while manual evaluation does not scale to the volume and diversity of production outputs \cite{liu2023itemresponse}.

Recent work shows that LLMs can approximate human judgments in relevance assessment and information retrieval tasks \cite{zheng2023judge,liu2023geval,faggioli2023perspectives,thomas2024preferences}. However, most such systems are designed for offline labeling or fixed evaluation sets. They do not directly address the needs of continuous monitoring: targeted sampling, model-specific thresholding, structured alerting, failure diagnosis, and operational triage in live recommender systems.

We present Model Sanity Service (MOSS), a production framework for monitoring and diagnosing semantic quality in product discovery systems. MOSS integrates LLM-based evaluation into a full monitoring loop that supports recommendations, search, and offers. A key feature is a two-stage design: the first stage scores relevance, while a follow-up prompt categorizes low-scoring cases into failure modes such as category mismatch, intent mismatch, and attribute inconsistency. This turns isolated scores into reviewer-facing diagnostics and aggregate root-cause signals.

\subsection{Contributions}
We make four contributions:
\begin{enumerate}
  \item We introduce MOSS, a production framework for continuous semantic-quality monitoring that combines sampling, task-aware LLM scoring, structured alerting, and log-based analysis.
  \item We propose a two-stage evaluation design that extends scoring with structured diagnostic prompting, enabling systematic failure-mode analysis and triage.
  \item We evaluate MOSS on a 498-pair human-labeled recommendation benchmark against two non-LLM baselines (internal production embeddings and SBERT), an offer relevance study (113 pairs, 83.2\% confirmation rate), and a 466-pair search benchmark spanning 21 queries, showing that a single judge, score scale, and alert threshold can support cross-surface monitoring.
  \item We extend MOSS with a lightweight per-strategy drift-detection module that maintains score and pair-level alert-rate histories and independently flags abrupt and gradual changes in the judge or upstream production system.
\end{enumerate}

\section{Related Work}
Recommendation quality in practice is often assessed through aggregate metrics such as CTR, conversion, and revenue, while semantic relevance relies on human review. Prior work has studied recommender evaluation limitations \cite{zangerle2022survey,liu2023itemresponse}, LLMs as judges \cite{zheng2023judge,liu2023geval}, and LLM-based relevance assessment in search and product discovery \cite{faggioli2023perspectives,thomas2024preferences,upadhyay2024umbrela,mehrdad2024productsearch}. Industrial systems face additional constraints around scale, serving complexity, calibration, and operational evaluation \cite{zou2025survey,wang2025pinterest}.

Several lines of work are adjacent to MOSS but address different problems. Recommender observability systems monitor behavioral metrics rather than semantic relevance. LLM-judge failure modes (position bias, verbosity bias, self-enhancement bias) have been documented \cite{zheng2023judge}; MOSS mitigates some by using structured numeric scores rather than pairwise preferences. Evaluator calibration and human-in-the-loop triage are well-studied but have not been applied to production recommender monitoring; MOSS's alert-then-review workflow is a lightweight instance of both patterns.

MOSS differs from offline labeling pipelines by embedding LLM relevance assessment into a recurring monitoring workflow with alert thresholds, diagnostic prompting, human triage, and production incident analysis. Recent work also notes that LLM-based recommender methods can exhibit biases such as popularity bias \cite{lichtenberg2024popularity}; this motivates treating MOSS alerts as triage signals rather than ground truth.

\section{MOSS System Design and Workflow}
\subsection{Overview}
\begin{figure}[t]
\centering
\includegraphics[width=\columnwidth]{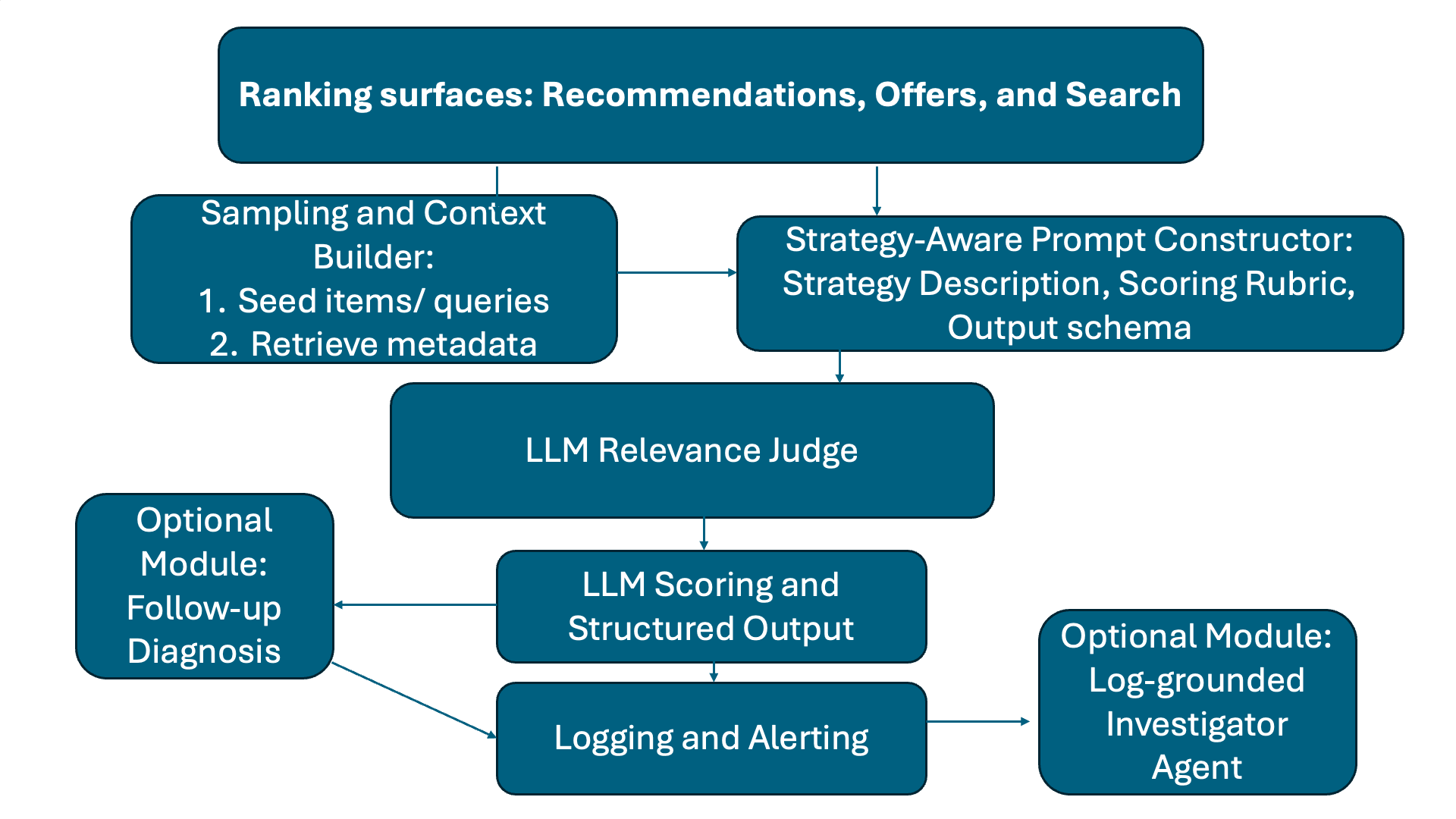}
\caption{MOSS system design overview.}
\label{fig:moss-overview}
\end{figure}

MOSS monitors semantic quality of production ranking outputs. In each run, it samples representative contexts, retrieves the corresponding ranked results, and uses an LLM with task-specific prompts to generate structured relevance scores and explanations. The resulting judgments are aggregated into monitoring metrics and alerts, while detailed logs remain available for inspection and recurring-pattern analysis. Figure~\ref{fig:moss-overview} summarizes the workflow. Table~\ref{tab:product-use-cases} previews three anonymized examples of the types of failures MOSS surfaces in production; the evaluation methodology and quantitative results follow in Sections~\ref{sec:eval}--\ref{sec:prod}.

\begin{table*}[t]
\caption{Anonymized recommendation failures surfaced by MOSS, illustrating distinct failure modes.}
\label{tab:product-use-cases}
\centering
\footnotesize
\setlength{\tabcolsep}{4pt}
\begin{tabular}{P{0.21\textwidth} P{0.21\textwidth} P{0.48\textwidth}}
\toprule
Seed item & Low-scored outputs & MOSS judgment \\
\midrule
BRAND\_A SIZE\_A PRODUCT\_NAME\_A (specific type of flavored beer, 6pk) & Diapers; oven cleaner & Scores: all 1/10. Items are unrelated household or party supplies rather than beverages. \\
BRAND\_D SIZE\_D PRODUCT\_NAME\_D (artificial Christmas tree) & Halloween decoration; toy-car gift wrap & Scores: 2/10 and 1/10. Items are off-season or off-theme for Christmas tree decor. \\
BRAND\_G SIZE\_G PRODUCT\_NAME\_G (pacifier gift set for baby) & Teen fiction novel; adult skincare set; board game for ages 12+ (among 10 items with scores $<5$) & Scores: all $\leq 4/10$ across the 10 flagged outputs. Failure modes span category mismatches (non-baby products), intent mismatches with infant-care essentials, and attribute mismatches around age group and product function. \\
\bottomrule
\end{tabular}
\end{table*}

\subsection{Sampling, Prompting, and Alerting}
MOSS uses weighted sampling to prioritize high-impact items while maintaining catalog coverage; weights are proportional to item traffic volume with a floor to ensure long-tail representation. Because sampling is non-uniform, aggregate metrics from production runs (e.g., mean score, alert rate) reflect the weighted distribution rather than unbiased catalog-level estimates. Sampled items are identified by product title; richer metadata (brand, price, description, and structured attributes such as color, material, style, and target audience) is available in MOSS logs and used for diagnostic review, but the scoring prompts reported here use item titles plus strategy context to match the deployed monitoring configuration. Recommendation prompts include the seed item title, candidate result title, and strategy-specific task description, such as complementary bundling, substitution, or related-item retrieval. Search prompts use the same structure over query-result pairs. Offers prompts follow the same structure over item-offer pairs.

LLM responses are required to follow a machine-parsable structure with a numerical score and concise reasoning. Parsed scores support aggregate metrics such as mean score, median score, alert rate, and threshold-based degradation signals. Alerts include the task type, item identifiers, score, and reasoning; both alerted and non-alerted cases are logged for auditability and downstream analysis.

\subsection{Diagnostic Prompting}
When an initial relevance score falls below a configurable threshold, MOSS triggers a second-stage diagnostic prompt. This prompt receives the original context (strategy description, seed item, recommended item), the initial score, and the judge's reasoning. The LLM is instructed to classify the failure into predefined categories (category mismatch, attribute mismatch, intent mismatch, ranking strategy issue, data quality issue, or evaluation model limitation) and returns a structured JSON response containing a primary cause, optional secondary causes, a brief explanation, and a suggested next step. This second stage transforms individual low scores into structured failure-mode labels that support aggregate root-cause analysis across strategies, categories, and serving contexts.

\subsection{Log-Grounded Investigator Agent}
\label{subsec:investigator}
To complement our automated scoring pipeline, we introduce an interactive AI agent that enables users to inspect individual LLM-based evaluations and uncover broader patterns across score distributions. While the scoring framework provides standardized relevance scores and justifications at scale, practitioners often need to understand why certain recommendations underperform and what aggregate trends may indicate misalignment. The AI agent supports this exploratory analysis through a structured yet flexible dialogue interface.

\begin{figure}[t]
\centering
\includegraphics[width=\columnwidth]{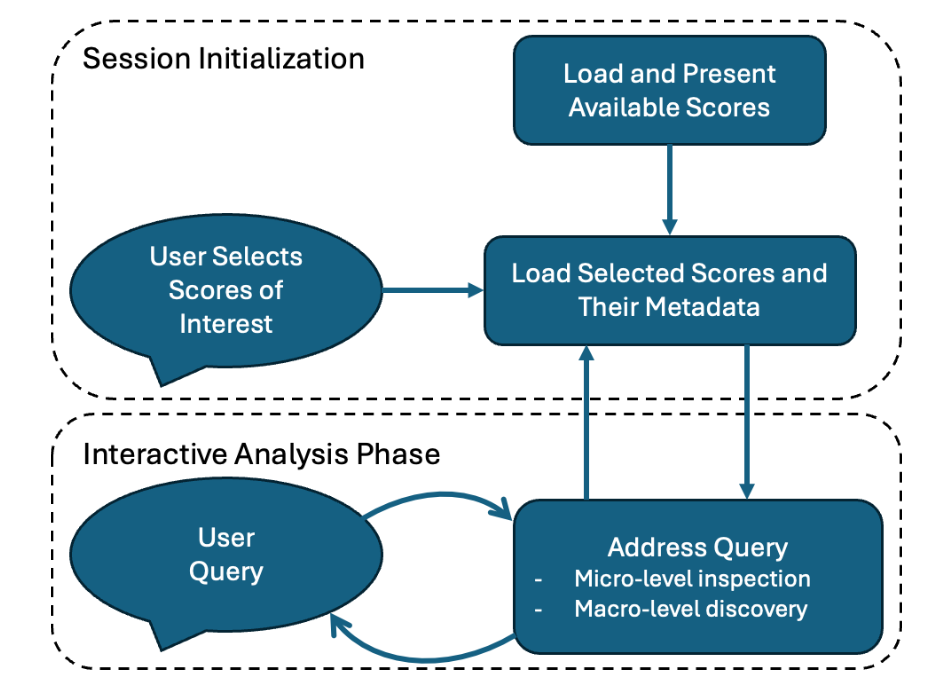}
\caption{Interactive Agent Workflow.}
\label{fig:investigator-workflow}
\end{figure}

The agent maintains conversational state and orchestrates the retrieval and interpretation of evaluation scores and metadata, as shown in Figure~\ref{fig:investigator-workflow}. After loading a selected set of scores and their associated metadata (e.g., product details, strategy identifiers, generated justifications), it enters an interactive analysis phase in which users issue free-form analytical queries. The AI agent supports two complementary modes:

\begin{itemize}
  \item \textbf{Micro-level inspection of individual evaluations.} Users can request explanations for specific high or low performing scores, surfacing the LLM's rationale alongside relevant metadata to diagnose issues such as misinterpreted attributes, intent mismatches, or brittle strategies.
  \item \textbf{Macro-level pattern discovery and trend analysis.} The agent can aggregate scores across strategies or user-defined subsets, enabling questions such as ``Are there common patterns among the lowest-scoring evaluations?'' or ``Which outliers exist within this strategy?'' By interpreting the scores and their accompanying details through LLM-augmented reasoning, the agent reveals trends that would otherwise require manual slicing or external analytical tooling.
\end{itemize}

By adding a conversational analytical layer to our evaluation framework, the agent transforms static score outputs into interactive analytical artifacts. It enables users across technical backgrounds to diagnose misaligned recommendations, validate strong ones, and understand system behavior at both granular and holistic levels. This contributes to the broader goal of making recommendation systems more transparent, explainable, and aligned with user and business intent.

\subsection{Prompt Structure and Inference Settings}
\label{subsec:prompt}
The scoring prompt provides the judge with (1)~the strategy name and a multi-paragraph strategy description defining the recommendation goal and scoring rubric, (2)~the seed item title, and (3)~the list of recommended item titles. For offers, the prompt provides the item context and the offer text, asking the judge to score contextual relevance. For search, the prompt replaces the seed item with the query and the candidate list with query results, asking the judge to score whether each result satisfies the shopper intent. The judge outputs a JSON array of integer scores (1--10) followed by free-text reasoning. All LLM scoring uses temperature 0.3 via an internal enterprise API endpoint. The production-deployed judge is o3, using a calibrated prompt variant that modifies the response format and scoring instructions to improve operational parseability and reduce alert volume (Section~\ref{subsec:strategy}); benchmark comparisons additionally use GPT-5, Gemini~2.5~Pro, and Mixtral~8$\times$7B under the original prompt template and temperature. Malformed or failed API responses are retried up to three times with exponential backoff.

\section{Analysis and Evaluation}
\label{sec:eval}
\subsection{Human Annotation Protocol and Baselines}
To evaluate the effectiveness of MOSS on item recommendations, we constructed a human-annotated benchmark specifically for this surface; separate evaluation datasets for offers and search are described in Section~\ref{sec:prod}. The recommendation benchmark consists of 498 seed-candidate recommendation pairs, evenly distributed across three recommendation strategies: Related (items from the same or adjacent categories), Bundles (complementary items typically purchased together), and Substitutes (functionally equivalent alternatives). The benchmark contains 166 pairs per strategy (before production-embedding coverage filtering). Alert-worthy prevalence differs by strategy: 27.1\% for Related, 13.3\% for Bundles, and 12.0\% for Substitutes (17.5\% overall).

\textbf{Annotation procedure.} Two domain-expert reviewers independently labeled each pair using full product metadata and the strategy definition (Related, Bundles, or Substitutes). Reviewers were blinded to all model scores. The rubric defined an alert-worthy pair as one where the recommended item would be poor or irrelevant for the given strategy; plausible recommendations were labeled non-alert-worthy. Borderline cases were the main source of disagreement and were adjudicated by a third expert reviewer using the same metadata and strategy context.

\textbf{Inter-annotator agreement.} The reviewers achieved 85.34\% raw agreement with Cohen's $\kappa=0.4894$ \cite{cohen1960agreement}. The moderate $\kappa$ reflects high chance agreement (71.1\%) due to 17.5\% prevalence; prevalence-adjusted kappa (PABAK) is 0.707 (``substantial agreement''). The final benchmark contains 411 non-alert-worthy and 87 alert-worthy pairs. o3 agreement with individual reviewers (84.5\% and 84.3\%, $\kappa=0.523$ and $0.521$) is similar to inter-human agreement.

\textbf{Non-LLM baselines.} We compare LLM judges with two non-LLM baselines. The SBERT baseline represents each item using a rich text string containing title, brand, description, and structured attributes, encoded with \texttt{all-mpnet-base-v2} \cite{reimers2019sentencebert}; cosine similarity is thresholded to flag low-relevance pairs. The production-embedding baseline uses an internal production embedding signal trained from aggregate behavioral co-occurrence and catalog taxonomy signals; it is the strongest non-LLM signal available in production. Production embeddings cover 421 of the 498 benchmark pairs (84.5\%), so we use this 421-pair coverage subset as the primary matched comparison for all methods. The covered subset contains 77 alert-worthy pairs out of 421 (18.3\% prevalence) and the uncovered subset contains 10 alert-worthy pairs out of 77 (13.0\%); together these account for the full 87 alert-worthy pairs in the benchmark. The similar prevalence across the two subsets suggests no major selection bias from the coverage restriction.

An important design distinction: LLM judges receive strategy context (e.g., ``complementary bundle'') while embedding baselines are strategy-agnostic. Conversely, SBERT uses enriched item text (title, brand, description, attributes) whereas LLM prompts use titles only, matching the deployed configuration. These asymmetries work in opposite directions; the comparison reflects the practical deployment choice between a general-purpose similarity signal and a task-aware LLM evaluator, not a controlled architecture ablation.

\subsection{Comparative Model Performance}
\begin{figure}[htbp]
\centering
\includegraphics[width=\columnwidth]{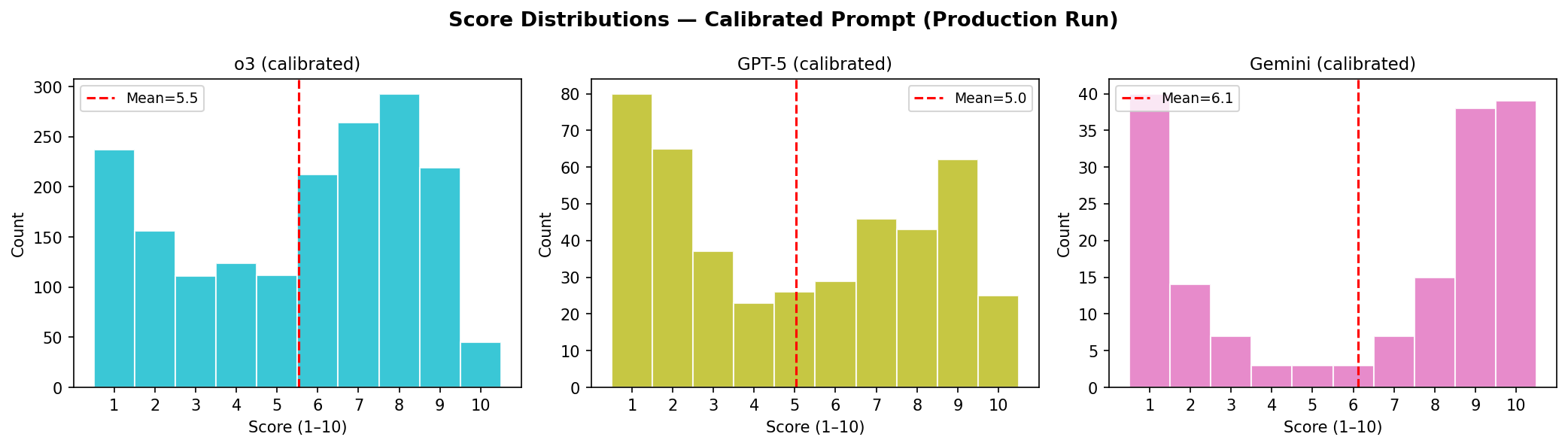}
\caption{Recommendation score distributions by LLM judge from a diagnostic run using the o3-calibrated prompt. Different judges use the 1--10 scale differently under the same scoring guidance, motivating judge-specific calibration. Production deployment uses the calibrated o3 configuration.}
\label{fig:judge-score-distribution}
\end{figure}

\begin{figure}[t]
\centering
\includegraphics[width=\columnwidth]{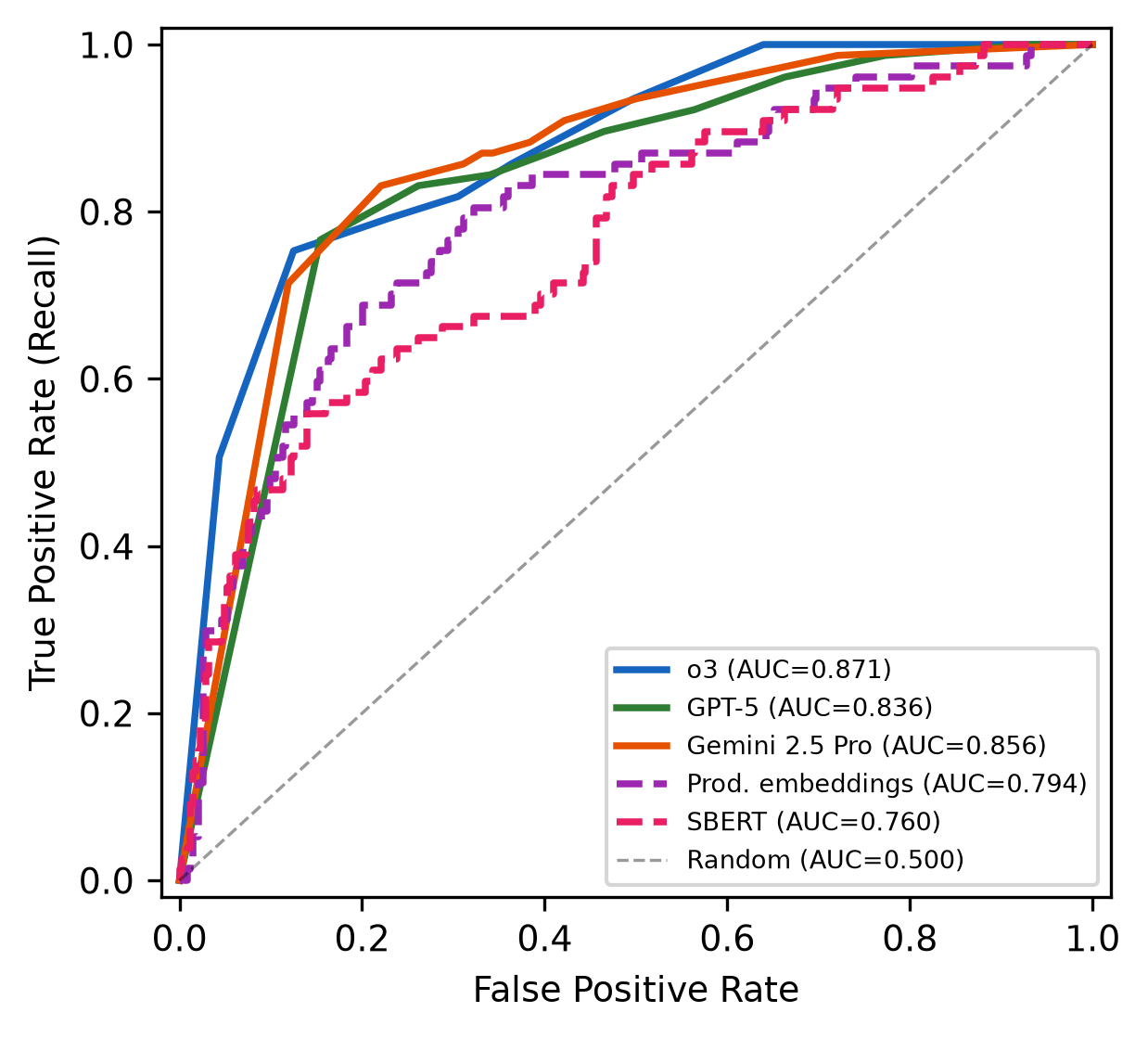}
\caption{ROC comparison between LLM judges and non-LLM baselines on the 421-pair coverage matched subset. AUC estimates and 95\% bootstrap confidence intervals are reported in Table~\ref{tab:model-performance}.}
\label{fig:llm-vs-baselines-roc}
\end{figure}

We evaluate three frontier proprietary judges (o3, GPT-5, and Gemini 2.5 Pro) on the 421-pair coverage subset. Figure~\ref{fig:judge-score-distribution} shows their score distributions under the calibrated prompt described in Section~\ref{subsec:prompt}, illustrating that different judges use the 1--10 scale differently under identical scoring guidance and motivating model-specific thresholding rather than a universal cutoff. Table~\ref{tab:model-performance} additionally includes Mixtral 8$\times$7B as an earlier open model for completeness, evaluated on its own coverage-matched subset (Section~\ref{subsec:prompt}); it is excluded from Figure~\ref{fig:llm-vs-baselines-roc} because it does not reach full 421-pair coverage. Thresholds are selected retrospectively by maximizing F1 on the evaluation set; Section~\ref{subsec:loso} presents a leave-one-strategy-out analysis that directly tests threshold generalization across strategies. Table~\ref{tab:model-performance} reports precision, recall, F1, false-positive rate (FPR), and AUC with 95\% bootstrap confidence intervals.

The calibrated prompt is deployed despite lower benchmark F1 and AUC because it optimizes the operational alerting objective rather than aggregate benchmark quality. We therefore separate benchmark ranking quality from the production deployment choice. The original prompt provides the strongest benchmark ranking quality (F1 and AUC), and we use it as the reference configuration for cross-model comparisons. The calibrated prompt was selected for production specifically for the o3 judge and does not transfer to other models (Section~\ref{subsec:strategy}); it was chosen for three operational reasons: it improves precision and reduces FPR on Related (the highest-volume strategy in our monitoring loop, Section~\ref{subsec:strategy}); it produces a more reliably machine-parsable response format that reduces parsing failures and retries; and it lowers the production alert volume into a range that the human triage queue can sustain. The benchmark cost is a lower F1 and AUC, which we report transparently in Table~\ref{tab:model-performance}.

\begin{table*}[t]
\caption{Performance on the 421-pair coverage matched subset (F1-maximized thresholds, bootstrap 95\% CIs, 2\,000 resamples). The calibrated o3 row is the production-deployed configuration (N=419 due to two missing scores); Mixtral 8$\times$7B is evaluated on N=397 pairs due to 24 missing judge outputs; remaining LLM rows use the original prompt (N=421). See Section~\ref{subsec:prompt} for prompt details.}

\label{tab:model-performance}
\centering
\small
\setlength{\tabcolsep}{3pt}
\begin{tabular}{lccccccc}
\toprule
Model & N & T & Prec. & Rec. & F1 [95\% CI] & FPR & AUC [95\% CI] \\
\midrule
\multicolumn{8}{l}{\textit{LLM judges}} \\
o3 (original) & 421 & 3 & 0.574 & 0.753 & 0.652 [0.561, 0.726] & 0.126 & 0.871 [0.827, 0.911] \\
o3 (calibrated, deployed) & 419 & 3 & 0.529 & 0.584 & 0.556 [0.458, 0.646] & 0.117 & 0.829 [0.779, 0.879] \\
GPT-5 & 421 & 2 & 0.527 & 0.766 & 0.624 [0.535, 0.698] & 0.154 & 0.836 [0.785, 0.879] \\
Gemini 2.5 Pro & 421 & 2 & 0.573 & 0.714 & 0.636 [0.543, 0.710] & 0.119 & 0.856 [0.809, 0.898] \\
Mixtral 8$\times$7B & 397 & 5 & 0.380 & 0.721 & 0.497 [0.409, 0.577] & 0.243 & 0.769 [0.707, 0.826] \\
\midrule
\multicolumn{8}{l}{\textit{Non-LLM baselines}} \\
Prod. embeddings & 421 & 0.603 & 0.458 & 0.636 & 0.533 [0.437, 0.611] & 0.169 & 0.794 [0.733, 0.850] \\
SBERT (mpnet + enriched) & 421 & 0.264 & 0.462 & 0.558 & 0.506 [0.405, 0.592] & 0.145 & 0.760 [0.694, 0.816] \\
\bottomrule
\end{tabular}
\end{table*}

Table~\ref{tab:model-performance} reports all complete-coverage models on the 421-pair coverage subset; calibrated o3 is evaluated on 419 pairs due to two missing scores, and Mixtral 8$\times$7B is evaluated on 397 pairs due to 24 missing judge outputs. Original-prompt o3 achieves the highest F1 (0.652) and AUC (0.871); the calibrated o3 prompt deployed in production trades recall for lower FPR (from 0.126 to 0.117) and alert volume (from 24.1\% to 20.3\%), at the cost of F1 (from 0.652 to 0.556) and AUC (from 0.871 to 0.829). All original-prompt frontier LLM judges show higher AUC point estimates than both embedding baselines; the deployed calibrated o3 configuration also remains above both embedding baselines, but with lower benchmark F1/AUC than original-prompt o3. Because marginal bootstrap intervals overlap in some comparisons, we avoid claiming formal statistical significance and leave paired AUC-difference testing to future work. Figure~\ref{fig:llm-vs-baselines-roc} provides the matched-subset ROC comparison.

The calibrated and original o3 prompts use identical strategy descriptions, but the calibrated prompt changes the response format and scoring instructions. Although intended as an operational parsing and alert-quality improvement, this prompt change affected score distributions and ranking quality heterogeneously across strategies (Section~\ref{subsec:strategy}). Bootstrap CIs in Table~\ref{tab:model-performance} fix the threshold at the full-data optimum within each resample; the LOSO analysis (Section~\ref{subsec:loso}) provides the more conservative threshold-generalization estimate for original-prompt judges.

Prompt calibration did not transfer uniformly across judges: applying the o3-calibrated guidance to GPT-5 and Gemini degraded AUC from 0.836 to 0.528 and from 0.856 to 0.668, respectively. We therefore use the original common prompt for cross-model comparison and report the calibrated prompt only for o3, highlighting that judge calibration is model-specific.

\subsubsection{Threshold Generalization: Leave-One-Strategy-Out Evaluation}
\label{subsec:loso}
Because thresholds in Table~\ref{tab:model-performance} are selected by maximizing F1 on the 421-pair matched subset, they represent in-sample operating-point estimates rather than held-out performance. We evaluate threshold transfer for original-prompt judges using leave-one-strategy-out (LOSO) evaluation on the same 421-pair subset: for each of the three strategies (Related, Bundles, Substitutes), we tune the threshold on the other two strategies and evaluate on the held-out one. This tests both threshold stability and cross-strategy generalization. We report the LOSO results as a macro-average over the three held-out folds (i.e., F1 is computed separately on each held-out strategy and the three per-strategy F1 values are averaged with equal weight, so each strategy contributes equally regardless of its pair count).

Table~\ref{tab:loso} reports the LOSO macro-averaged results. For all three frontier judges, the LOSO-selected threshold exactly matches the full-set optimum, and LOSO macro-F1 is within 0.01 of the full-set F1. This indicates that the F1-maximized thresholds in Table~\ref{tab:model-performance} are not overfit to strategy composition.

\begin{table}[t]
\caption{LOSO threshold generalization for original-prompt judges on the 421-pair coverage subset (macro-average over 3 strategy folds, defined in Section~\ref{subsec:loso}). ``FS-F1'' = full-set F1 at LOSO~T; ``Opt-F1'' = full-set F1-maximized optimum.}
\label{tab:loso}
\centering
\scriptsize
\setlength{\tabcolsep}{2pt}
\begin{tabular}{lcccccc}
\toprule
Model & T & Prec & Rec & F1 & FS-F1 & Opt-F1 \\
\midrule
o3        & 3   & 0.616 & 0.739 & 0.648 & 0.652 & 0.652 \\
GPT-5     & 2   & 0.554 & 0.756 & 0.622 & 0.624 & 0.624 \\
Gemini    & 2   & 0.623 & 0.713 & 0.633 & 0.636 & 0.636 \\
\bottomrule
\end{tabular}
\end{table}

The LOSO analysis addresses the threshold-selection leakage concern: for all original-prompt frontier judges, the LOSO-selected threshold matches the full-set optimum exactly and performance generalizes with only marginal degradation.

\subsection{Strategy-Level Effects and Prompt Calibration}
\label{subsec:strategy}
Table~\ref{tab:strategy-performance} shows that the LLM advantage is not uniform across recommendation strategies. Production embeddings are competitive for Related items, where category and attribute overlap are useful. The gap is largest for Bundles and Substitutes, where relevant items may be intentionally dissimilar or functionally equivalent and therefore require relational reasoning beyond cosine similarity. To separate threshold-selection effects from model quality, the table uses the global threshold from Table~\ref{tab:model-performance} for each judge. Per-strategy F1-maximized (oracle) thresholds improve performance (e.g., o3 Related F1 rises from 0.640 to 0.729 at $T=2$, and Bundles from 0.395 to 0.508), confirming that a single global threshold is a deployment compromise. Per-strategy thresholds are a natural production extension.

\begin{table}[t]
\caption{Per-strategy performance on the 421-pair coverage subset using global thresholds from Table~\ref{tab:model-performance} (original prompt).}
\label{tab:strategy-performance}
\centering
\small
\setlength{\tabcolsep}{2pt}
\begin{tabular}{llccccc}
\toprule
Strategy & Model & T & Prec. & Rec. & F1 & FPR \\
\midrule
Related & Prod. emb. & 0.603 & 0.627 & 0.821 & 0.711 & 0.207 \\
 & SBERT & 0.264 & 0.684 & 0.667 & 0.675 & 0.130 \\
 & o3 & 3 & 0.525 & 0.821 & 0.640 & 0.315 \\
 & GPT-5 & 2 & 0.571 & 0.821 & 0.674 & 0.261 \\
 & Gemini & 2 & 0.727 & 0.821 & 0.771 & 0.130 \\
\midrule
Bundles & Prod. emb. & 0.603 & 0.250 & 0.421 & 0.314 & 0.188 \\
 & SBERT & 0.264 & 0.256 & 0.526 & 0.345 & 0.227 \\
 & o3 & 3 & 0.263 & 0.789 & 0.395 & 0.328 \\
 & GPT-5 & 2 & 0.239 & 0.842 & 0.372 & 0.398 \\
 & Gemini & 2 & 0.250 & 0.895 & 0.391 & 0.398 \\
\midrule
Subs. & Prod. emb. & 0.603 & 0.375 & 0.474 & 0.419 & 0.121 \\
 & SBERT & 0.264 & 0.438 & 0.368 & 0.400 & 0.073 \\
 & o3 & 3 & 0.636 & 0.737 & 0.683 & 0.065 \\
 & GPT-5 & 2 & 0.516 & 0.842 & 0.640 & 0.121 \\
 & Gemini & 2 & 0.536 & 0.789 & 0.638 & 0.105 \\
\bottomrule
\end{tabular}
\end{table}

A key deployment challenge is alert fatigue. The calibrated prompt preserves the same strategy descriptions but changes the response format and scoring instructions to improve operational parseability. On a matched 419-pair subset where both prompts produced valid scores, the effect is heterogeneous across strategies: for Related (the highest-volume production strategy), calibration improved precision (from 0.674 to 0.757) and reduced FPR (from 0.163 to 0.098) while maintaining F1 (from 0.729 to 0.737). For Substitutes, calibration degraded both recall (from 0.632 to 0.316) and AUC (from 0.912 to 0.778), indicating that the prompt change altered how the judge reasons about functional equivalence. Bundles showed mild degradation (F1 from 0.508 to 0.415).

Together, these results indicate that prompt calibration should be optimized against the operational alerting objective, not only aggregate benchmark F1. Per-strategy and per-judge prompt calibration is planned as future work.

\begin{figure}[t]
\centering
\includegraphics[width=\columnwidth]{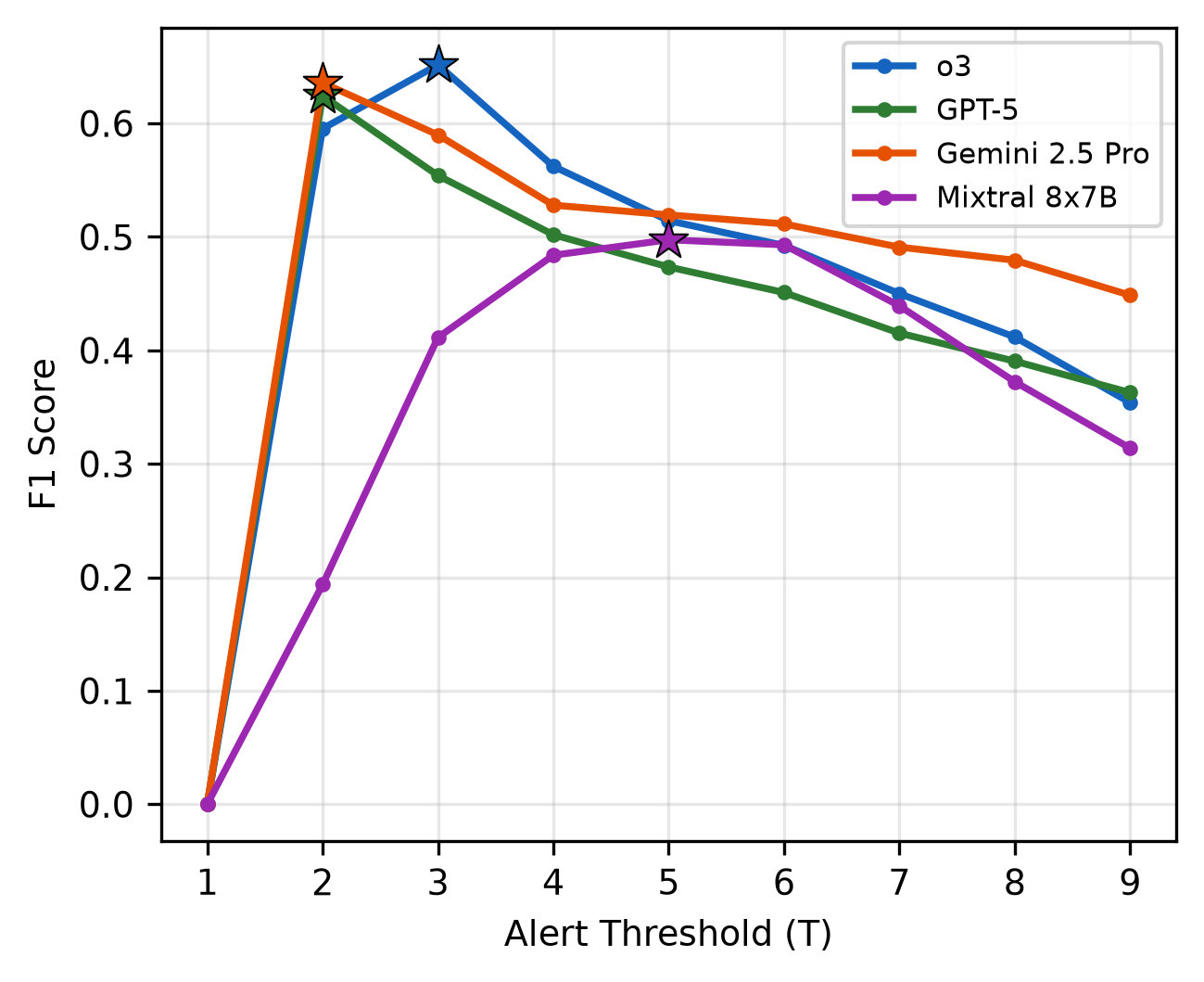}
\caption{Retrospective threshold sensitivity on the human-labeled evaluation set (original prompt). Operating points differ by judge, reinforcing that threshold selection is a deployment parameter.}
\label{fig:threshold-sensitivity}
\end{figure}

In a production monitoring run with the calibrated prompt, o3 scored 1,773 unique seed--recommendation pairs across the three strategies. Table~\ref{tab:cal-production-summary} summarizes score statistics and alert rates. The overall production alert rate (28.4\%) exceeds the matched-benchmark calibrated alert rate (20.3\% at $T\leq3$), reflecting broader catalog coverage, long-tail sampling, and different production candidate distributions.

\begin{table}[htbp]
\caption{Calibrated o3 production run: score statistics and alert rates by strategy (1,773 pairs, threshold $\leq3$).}
\label{tab:cal-production-summary}
\centering
\small
\setlength{\tabcolsep}{4pt}
\begin{tabular}{lcccc}
\toprule
Strategy & N & Mean & Median & Alert rate \\
\midrule
Related & 1,065 & 5.26 & 6.0 & 32.8\% \\
Substitutes & 461 & 6.54 & 7.0 & 13.0\% \\
Bundles & 247 & 4.88 & 5.0 & 38.5\% \\
All & 1,773 & 5.54 & 6.0 & 28.4\% \\
\bottomrule
\end{tabular}

\end{table}

\begin{figure}[htb]
\centering
\includegraphics[width=\columnwidth]{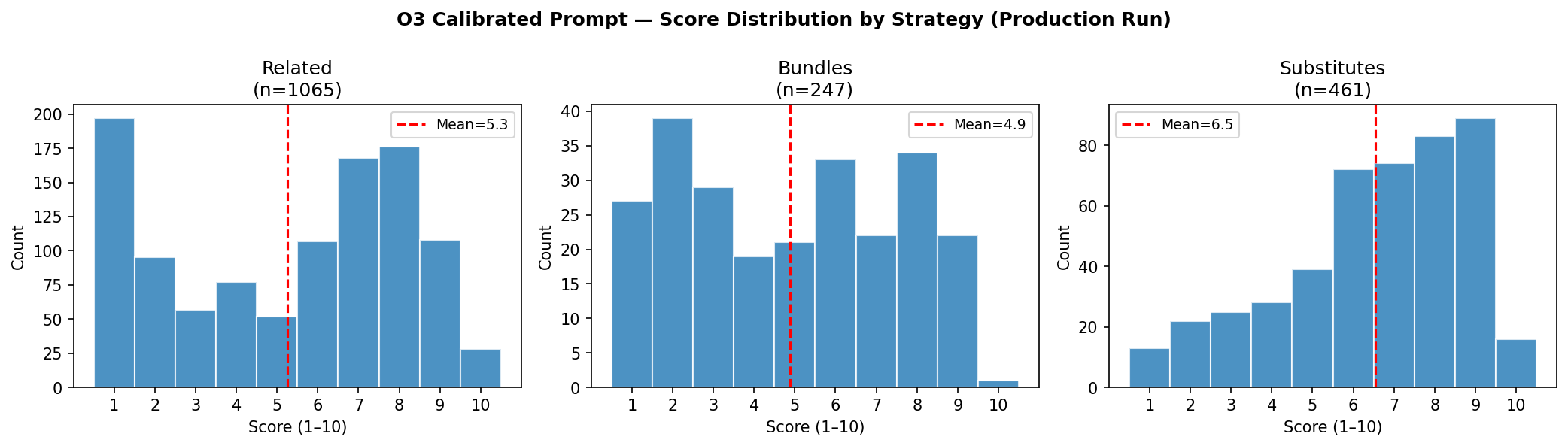}
\caption{Calibrated o3 score distribution by recommendation strategy in the 1,773-pair production monitoring run. Strategy-specific distributions explain why alert rates should be interpreted relative to task intent.}
\label{fig:cal-score-by-strategy}
\end{figure}

\subsection{Follow-up Triage and Operational Utility}
To evaluate the operational usefulness of the two-stage alerting workflow, we conducted an expert triage study before the calibrated prompt was deployed. The study reviewed 116 alerts produced by the original o3 prompt at threshold $T=2$ in a pre-calibration benchmark run. Experts confirmed 98 as actionable (84.5\%; Wilson 95\% CI: 76.8--90.0\%), with actionability rates of 78.3\% for Bundles, 84.9\% for Related, and 100\% for Substitutes. Of these 116 alerts, 65 met the strict semantic-mismatch benchmark definition (benchmark precision 0.560), while 98 were judged operationally actionable under a broader triage rubric that includes borderline cases warranting human review. The gap reflects the difference between a binary mismatch label and a production reviewer's judgment about whether a case merits investigation; reviewers noted that strategy definitions, particularly for Bundles, are inherently ambiguous, and borderline pairs were frequently judged worth reviewing even when they did not meet the strict mismatch criterion.

\begin{table}[htbp]
\caption{Expert triage of 116 second-stage alerts (o3 original prompt, $T=2$).}
\label{tab:expert-triage}
\centering
\small
\setlength{\tabcolsep}{4pt}
\begin{tabular}{lcccc}
\toprule
Task & Reviewed & Actionable & Wrong/unclear & Rate \\
\midrule
Bundles & 46 & 36 & 10 & 78.3\% \\
Related & 53 & 45 & 8 & 84.9\% \\
Substitutes & 17 & 17 & 0 & 100.0\% \\
Total & 116 & 98 & 18 & 84.5\% \\
\bottomrule
\end{tabular}
\end{table}

\section{Drift Detection for the Monitoring Loop}
\label{sec:drift}
Two failure modes can pose long-term risks to MOSS. First, the proprietary judge can change silently between runs; the model behind an enterprise endpoint may be changed without notice. Section~\ref{sec:eval} shows that a comparable-magnitude shift in prompt or response format moves AUC by roughly 0.04 on o3 and materially degrades GPT-5 and Gemini under the same guidance. Second, per-strategy scores can decay very gradually due to changes to upstream data or user behavior which impact model performance. To guard against these failure modes, we extend MOSS with a drift-detection module which runs alongside the monitoring loop.

\subsection{Design}
To detect drift, every MOSS run appends one JSONL record per strategy to a persistent history file. Each record in the file captures the run identifier, timestamp, judge identity, strategy, per-strategy mean score, per-strategy median score, pair-level alert rate (defined as the percentage of seed--candidate pairs scoring at or below the alert threshold), and pair count. At the end of each run, the drift check reads the history, computes per-strategy baselines over two windows, and compares the current run against them.

\subsection{Abrupt and Gradual Drift Checks}
The abrupt drift check compares the current run's mean score and pair-level alert rate against a 7-day weighted baseline (weights proportional to per-run pair count). At the 4-hour cadence, this window covers roughly 42 runs, which helps to smooth outliers while still catching regressions expeditiously. A score drop of at least 1.5 points or an alert-rate rise of at least 15 percentage points (pp) triggers a warning, while 2.5 points or 25 pp triggers a critical alert. A rise of 2.0 points also fires a ``suspicious rise'' alert, since abrupt upward moves in a stable monitor are more likely a data issue than a genuine quality improvement.

The gradual drift check targets slow erosion by comparing the 7-day baseline against a 30-day reference baseline. The score gap (long-baseline mean minus recent-baseline mean) and the alert-rate gap (recent-baseline pair-level alert rate minus long-baseline pair-level alert rate) capture sustained decline even when no single run trips the abrupt thresholds. A 1.0-point score gap or 10 pp alert-rate gap fires a warning, while 2.0 points or 20 pp fires a critical alert. This check activates only once the 30-day window contains records outside the 7-day window, in practice about a week after deployment. The two checks run independently per strategy per run and can both fire together.

\subsection{Operational Properties}
The drift checks add only a few reads and one append per run and are treated as non-critical: exceptions are logged but never fail the run. Alerts are routed to the same channel as per-item alerts, formatted with the baseline window size, the current-run value, the delta, and the name of the check which fired. A minimum-history guard suppresses drift alerts until the oldest baseline record is at least three days old, preventing noise after a new judge or strategy is onboarded. The design is intended to complement the periodic re-scoring of a fixed calibration set; the fixed set catches judge changes that leave production distributions untouched, while the history-based check catches changes that manifest in production between occasional calibration runs.

\section{Surface Evaluations and Production Use Cases}
\label{sec:prod}
\subsection{Product Recommendation Surface}
MOSS has run in production for over six months, monitoring three recommendation placements across three primary strategies (Related, Bundles, Substitutes) and two additional strategies on a recurring cadence of every four hours. Each run samples seed items, scores the corresponding recommendation candidates using the calibrated o3 prompt, and produces structured alert logs for human review. The 1,773-pair calibrated production run summarized in Table~\ref{tab:cal-production-summary} is one representative run; the monitoring loop has executed continuously across this period. Two production incidents illustrate the diagnostic value of the monitoring loop.

\textbf{Collaborative-filtering defect.} In a related-item placement, MOSS alerts were concentrated among long-tail items in a small set of categories rather than spread uniformly across the catalog. Investigation traced the pattern to an error in statistics calculations inside a collaborative-filtering step, degrading recommendations for a narrow but production-relevant slice of the catalog. Because the affected items represented a small share of total traffic, aggregate click or conversion metrics could easily dilute the signal. MOSS localized the issue to a specific serving context, category slice, and algorithmic component, enabling a targeted fix within the same monitoring cycle.

\textbf{Backfill blending defect (quantified).}
MOSS flagged high alert rates in one of our production bundle placements. For a children's board-book seed, the LLM judge assigned high scores to same-franchise items (10/10, 9/10) and moderate scores to adjacent children's products, but sharply penalized backfilled items intended for older audiences, including an adult horror paperback (1/10), an adult romance novel (1/10), and a teen fantasy novel (2/10). The score pattern localized the failure to the backfill source rather than the primary candidate source.

These alerts motivated efforts to improve the production bundling algorithms, leading to a newer enriched set of algorithms. To quantify the impact of these improvements, we ran a controlled offline experiment using MOSS on 99 matched seed items across the legacy and enriched placements. Table~\ref{tab:bundles-comparison} summarizes the results.

\begin{table}[t]
\caption{Paired MOSS comparison across legacy and enriched bundle placement configurations (99 shared seeds, calibrated o3, threshold $\leq3$, bootstrap 95\% CIs from 2{,}000 resamples).}
\label{tab:bundles-comparison}
\centering
\small
\setlength{\tabcolsep}{4pt}
\begin{tabular}{lrrrr}
\toprule
Placement & Seeds & Pairs & Mean score & Pair alert rate \\
\midrule
Legacy (live) & 99 & 995 & 4.68 & 41.6\% [35.5, 48.4] \\
Enriched (fix) & 99 & 348 & 6.93 & 9.8\% [5.4, 14.8] \\
\bottomrule
\end{tabular}
\end{table}

The legacy placement had a pair-level alert rate of 41.6\% [35.5, 48.4]. Switching to the enriched algorithms reduced the alert rate to 9.8\% [5.4, 14.8], a relative reduction of 76.4\% (absolute: $-$31.8 points, 95\% CI: [25.0, 39.0]). The two placements return candidate sets of different sizes by design, reflecting different candidate-generation strategies; the pair-count asymmetry is therefore an expected property of the comparison rather than a confound. This experiment shows that MOSS can not only surface a blending defect qualitatively but also quantify the improvement from a proposed fix before full production deployment. This ability to provide targeted, quantitative feedback on system health is a key requirement for a cross-surface monitoring framework; we now turn to MOSS's application on two other production surfaces.

\subsection{Offers and Search Surfaces}
Beyond recommendations, we apply MOSS to \emph{offers} (promotional content shown in shopping contexts) and \emph{search} (query--result relevance). Each surface uses a different evaluation design because the available ground-truth signal is different: the recommendation benchmark provides full human labels on a sampled set of pairs; for offers we adjudicate a reviewer-prioritized alert sample (an alert-enriched set, where the input is restricted to low-relevance candidates that MOSS already flagged) and report the rate at which alerts are confirmed as genuine issues; and the search surface is evaluated on a separate human-labeled benchmark of 466 query--result pairs across 21 product queries. Table~\ref{tab:cross-surface} summarizes the three settings so they can be compared at a glance.

\begin{table*}[t]
\caption{Cross-surface MOSS evaluation summary. ``Rate'' is alert-confirmation rate for alert-enriched sets and alert-worthy prevalence for human-labeled benchmarks. Search results (F1, AUC) are detailed in Table~\ref{tab:search-benchmark}.}
\label{tab:cross-surface}
\centering
\small
\setlength{\tabcolsep}{4pt}
\begin{tabular}{P{0.16\textwidth}ccP{0.11\textwidth}P{0.11\textwidth}P{0.34\textwidth}}
\toprule
Surface & N & Positives & Rate & Rate type & Scope \\
\midrule
Recs alerts & 116 & 98 & 84.5\% & confirmation & Second-stage follow-up alerts, original prompt, $T=2$ \\
Recs benchmark & 498 & 87 & 17.5\% & prevalence & Human-labeled benchmark; production-embedding comparison on 421-pair coverage subset \\
Offers & 113 & 94 & 83.2\% & confirmation & Alert-enriched review set with adjudication \\
Search benchmark & 466 & 186 & 39.9\% & prevalence & Human-labeled, 21 queries; o3 $T\leq3$; F1=0.831, AUC=0.884 \\
\bottomrule
\end{tabular}
\end{table*}

\textbf{Offers surface.} For the offers surface, we used an \emph{alert-enriched} evaluation design, scoring item-offer pairs using the offers prompt described in Section~\ref{subsec:prompt}. This approach, where reviewers adjudicate a sample of candidates that MOSS has already flagged as low-relevance, is a pragmatic choice when a full human-labeled benchmark is not available. It allows for rapid assessment of an alerting system's precision: if the system is flagging genuinely problematic content, the confirmation rate on its alerts will be high.

For our offers evaluation, 94 of 113 low-relevance candidates surfaced by MOSS were confirmed as genuinely problematic (83.2\%; Wilson 95\% CI: 75.2--89.0\%). The dominant failure modes were weak contextual ties (36.2\%), shopping-mission mismatches (34.0\%), and category mismatches (29.8\%). For example, offers for beauty, laundry, or diapers shown in the context of kids' award medals received very low relevance scores and were confirmed as poor contextual matches.

\textbf{Search surface.} For search, we constructed a full human-labeled benchmark of 466 query--result pairs across 21 anonymized product-intent queries drawn from an internal relevance-judgment dataset. The unit of evaluation is a single query--result pair: the query represents the shopper intent, and the result is one candidate product returned for that query. Raw query strings and system-specific retrieval details are not reported for confidentiality; instead, we report anonymized query identifiers and aggregate statistics. The selected queries cover low ($<$10\%), medium (10--35\%), and high ($>$49\%) alert-prevalence strata so that the benchmark tests both easy and failure-prone search contexts. Each pair was labeled alert-worthy (unrelated or attribute mismatch) or non-alert-worthy (strong or weak/related match), yielding 186 alert-worthy pairs (39.9\% prevalence).

The search experiment tests cross-surface transfer by re-using the judge, score scale, and alert threshold selected on the recommendation benchmark ($T{\leq}3$). This is a challenging test: if successful, it means MOSS can be extended to new surfaces with minimal tuning. Only the task context in the prompt is changed (see Section~\ref{subsec:prompt}): recommendation prompts ask whether a candidate item fits a seed item under a recommendation strategy, while the search prompt asks whether a candidate result satisfies the shopper intent expressed by the query. We scored all pairs with o3 at temperature 0.3 and applied the recommendation threshold without retuning on search. Table~\ref{tab:search-benchmark} reports results with 2{,}000-resample bootstrap 95\% CIs. In addition to AUC-ROC, we report PR-AUC (area under the precision-recall curve), which summarizes the tradeoff between precision and recall across all score thresholds. PR-AUC is more informative than ROC-AUC when the positive class is the operational target, as it is here: alert-worthy pairs make up 39.9\% of the benchmark, and a monitor's value depends on how precisely it identifies them.

\begin{table}[htbp]
\caption{Search benchmark results (N=466, 21 anonymized queries, o3, $T\leq3$, bootstrap 95\% CIs). F1-max threshold $T=5$ (F1=0.836). Selected per-query F1 is reported using anonymized query IDs; raw query strings are omitted for confidentiality.}
\label{tab:search-benchmark}
\centering
\small
\setlength{\tabcolsep}{4pt}
\begin{tabular}{lrrr}
\toprule
 & Prec. & Rec. & F1 [95\% CI] \\
\midrule
Overall & 0.762 & 0.914 & 0.831 [0.791, 0.868] \\
\quad AUC [95\% CI] & \multicolumn{3}{r}{0.884 [0.852, 0.911]} \\
\quad PR-AUC [95\% CI] & \multicolumn{3}{r}{0.742 [0.678, 0.804]} \\
\quad Macro-F1 (21 queries) & \multicolumn{3}{r}{0.597} \\
\midrule
\multicolumn{4}{l}{\textit{Selected per-query F1 (N, human alerts)}} \\
Query S1 (20, 9)  & \multicolumn{3}{r}{1.000} \\
Query S2 (20, 3)  & \multicolumn{3}{r}{1.000} \\
Query S3 (40, 32) & \multicolumn{3}{r}{0.952} \\
Query S4 (18, 8)  & \multicolumn{3}{r}{0.933} \\
Query S5 (85, 62) & \multicolumn{3}{r}{0.878} \\
Query S6 (20, 14) & \multicolumn{3}{r}{0.824} \\
\bottomrule
\end{tabular}
\end{table}

The transferred threshold $T\leq3$ achieves F1=0.831, within 0.005 of the search-optimal threshold $T=5$ (F1=0.836). As a lower-bound reference, the alert-all baseline (flagging every pair) yields F1=0.571 at 39.9\% prevalence; MOSS scores 0.260 F1 points above this baseline.

Table~\ref{tab:search-benchmark} reports both micro-F1 and macro-F1 because they answer different operational questions. Micro-F1 (0.831) pools all 466 query--result pairs together and computes a single F1 over the pooled decisions; it is the primary alerting metric because each candidate result contributes one review decision regardless of which query it came from. Macro-F1 (0.597) instead computes F1 separately for each of the 21 queries and then averages the per-query F1 scores with equal weight; it is included as a stress test of per-query variability, where any query with poor performance pulls the average down regardless of how many pairs it contains. Macro-F1 is lower because several low-prevalence queries have few or no human alerts; under our conservative convention, zero-positive queries contribute per-query F1=0 even when the model correctly produces few alerts. Overall, these results suggest that the judge, score scale, and recommendation-selected alert threshold can transfer to search when the prompt is adapted from item-to-item recommendation relevance to query--result relevance.

\section{Limitations, Ethics, and Privacy}
Several limitations remain. First, MOSS is a monitoring and triage layer, not an automated quality adjudicator: scores prioritize cases for review and diagnosis rather than replacing human judgment. LLM judges vary in calibration and can over-alert on borderline cases; generated rationales are therefore treated as review aids rather than authoritative explanations. Second, the evaluations differ in scope across surfaces because each surface uses a different evaluation design. Recommendations have full human labels and embedding baselines; offers use alert-enriched adjudication; search uses a 466-pair query--result benchmark. These differences support the intended monitoring use case but limit direct comparisons across surfaces.

\textbf{Evaluation scope.} Calibrated-prompt results and operating thresholds are retrospective estimates; we mitigate threshold-selection concerns with leave-one-strategy-out analysis, but a frozen-prompt prospective test remains future work. Bootstrap confidence intervals are marginal rather than paired, so we report point-estimate comparisons without formal significance claims. The non-LLM baselines (production embeddings and SBERT with enriched item text) represent the strongest strategy-agnostic signals available in production; the benchmark is sized for monitoring evaluation rather than for training supervised classifiers. A further asymmetry is that the deployed scoring prompts use item titles only, while annotators had access to full product metadata; where titles are vague or ambiguous, this can produce false alerts that a metadata-aware prompt would avoid, and quantifying that gap is future work.

\textbf{Production scope.} The production incidents provide diagnostic evidence, and the bundle case includes a quantified offline comparison; measuring downstream business impact through controlled online experiments remains future work. Production embeddings cover 421 of 498 benchmark pairs, so production-embedding comparisons apply only to that subset. The warning and critical thresholds are initial operator-selected heuristics chosen to balance sensitivity and alert fatigue; prospective evaluation of detection delay and false-alert rates remains future work.

\textbf{Privacy and governance.} MOSS operates on item, offer, query, and recommendation metadata rather than guest-level personal data; product examples are anonymized. All LLM scoring uses enterprise API endpoints governed by organizational data-use agreements. Search queries were sampled from high-volume head queries (product category terms). Prompt-injection filtering, query-level PII detection, and systematic fairness audits are future work. No guest behavioral records or identifiable user data were used.

\section{Conclusions}
Our results show that LLM-based relevance scoring, when embedded in a structured monitoring loop with sampling, thresholding, and diagnostic follow-up, can serve as a useful operational quality signal for production search and recommendation systems. The results support three operational observations for practitioners considering similar approaches.

First, strategy-aware LLM judges add the most value where embedding similarity falls short. On our recommendation benchmark, the advantage over production embeddings and SBERT baselines is concentrated on Bundles and Substitutes, tasks that require intent-level relational reasoning rather than surface-level item similarity. For Related items, where category and attribute overlap are strong cues, production embeddings remain competitive.

Second, a single judge, score scale, and threshold can transfer across surfaces. The o3 judge and alert threshold ($T{\leq}3$), selected on recommendations, achieve F1 0.831 and AUC 0.884 on a 466-pair search benchmark spanning 21 queries with diverse alert rates, within 0.005 of the search-optimal threshold, using only a surface-specific prompt adaptation. This substantially reduces the setup cost of extending monitoring to new surfaces.

Third, the operational value of MOSS lies in localization, not just detection. In production, MOSS surfaced two confirmed serving defects that aggregate behavioral metrics would likely miss. In the backfill blending case, a controlled offline comparison quantified a 76.4\% relative reduction in alert rate from the proposed fix, enabling data-driven prioritization before full deployment.

Future work will focus on per-strategy prompt calibration and held-out validation, frozen-prompt prospective evaluation to confirm threshold stability over time, controlled online experiments to measure downstream business impact, and operational instrumentation (detection latency, review cost, false-alert trends) to quantify the full monitoring loop.

\end{document}